# Enhancement of light transmission through randomly located copper nano-islands near the percolation threshold.

Luis G. Mendoza-Luna, Eva M. Rojas-Ruiz, José L. Hernández-Pozos*

Departamento de Física, Universidad Autónoma Metropolitana-Iztapalapa, Apartado Postal 55-534, 09340, Ciudad de México, México.

**ABSTRACT:** We report the absorption/transmission properties of thin copper films slightly below the percolation limit produced by laser ablation; these samples show randomly positioned islands. Two types of films are grown: in the first type, copper is deposited on glass, and it is shown that as the amount of deposited copper approaches the percolation limit, the samples show enhanced light transmission; in the second set of samples, made of $TiO_2$/Cu/$TiO_2$ layers, the presence of titanium oxide inhibits light transmission enhancement, even though the same amount of copper has been deposited compared to the samples deposited on glass. We also present numerical calculations based on the discrete dipole method to simulate the optical behaviour of the copper films. The enhanced light transmission of the samples presented in this work resembles the Extraordinary Optical Transmission (EOT) observed in periodic metallic structures. The similarities and differences between experiment and numerical calculations and the possible use of these films as sensors are discussed.

Thin metallic films or dielectric/metal/dielectric structures have been studied and used for many different purposes in the last few decades. Particularly, since the 1990's, as surface plasmon resonances in thin films and other nanostructures have been better understood, research work in first principles studies and applications has increased enormously.

Some examples of recent applications reported for this kind of systems are: Cu or Au films deposited on dielectric substrates for designing transparent electrodes or intelligent windows (1; 2; 3), plasmon-enhanced transparency and plasmon-enhanced photoluminescence of ZnO films with Ag or Au nanoislands (4; 5; 6; 7; 8; 9). In the cited examples, for metal deposition several techniques are used such as sputtering, thermal evaporation, chemical self-assembling reactions or electrochemical reactions. All these methods have in common that they produce a random array of nanoislands on the dielectric surface.

On the other hand, in 1998 it was observed that when a regular array of nano-sized holes etched in a metallic film was irradiated with visible/infrared light, the transmission of the sample was far larger than the one predicted by the usual electromagnetic treatment for small apertures (10); such behavior has been termed Extraordinary Optical Transmission (EOT) (11). In those experiments, silver films with thicknesses ranging between 200-500 nm were deposited on quartz substrates. The aperture array was drilled into the film with a Focused-Ion-Beam (FIB) system; the hole diameter was varied between 150-1000 nm and the periodicity of the apertures was changed between 600-1800 nm. Since those early studies, other research programs have been devoted to explaining the fine details of this phenomenon (see the reviews (12; 13)). The case of quasi-periodic structures showing transmission resonances has also been described in (14); in this work, a Fano-resonance-type mechanism is used to explain the enhancement of transmission. There is at least one report where a completely random array of holes is claimed to show EOT (15). Recently, several reports have appeared with sensors for gas or biomolecules based in periodic metallic nanostructures (e.g. (16; 17)).

It is certain that in the near future the use of sensors based on plasmonic-mediated phenomena will play an ever-increasing role in the applications of nanotechnology, either with random or periodic metallic structures. Thus, exploring diverse geometries and materials presenting enhanced transmission of light is likely to be paramount for the further development of these emerging technologies.

In this work, we demonstrate light transmission enhancement as the coverage of a Cu thin film deposited on glass increases towards the percolation threshold. Conversely, if a metal film is placed between two layers of dielectric of equal thickness, the enhancement of transmission is suppressed. We also show numerical simulations of light scattering by randomly positioned scatterers performed with the open-source software DDSCAT, which reproduce several features of the experiment.

## RESULTS AND DISCUSSION.

The samples produced for these experiments were manufactured by laser ablation. The following is a brief description of the experimental setup: samples were created by laser ablation of copper either with the second or third harmonic of a Nd:YAG laser in a UHV chamber maintained at pressures of around $10^{-5}$ mbar, and the distance between the copper target and the substrate was kept constant at 2.5 cm during all the experiments. The energy density of the beam was optimized to avoid, as much as possible, sputtering of the copper onto the glass substrates. The variations of the optical properties described in this paper are chiefly the result of varying the number of laser pulses (i.e. the quantity of deposited material) on the substrates. A full description of the experimental apparatus is given in the Methods section.

**Absorption of copper films below the percolation threshold deposited on glass.** Copper films were deposited on glass using fixed pulse energy of (4.0±0.3) mJ. Several samples series were fabricated with 1, 1.5, 2, 2.5, 3, 4 and 5 (x $10^3$) laser pulses; their absorbance was measured with a Cary 5E spectrophotometer (operating in the absorption mode).

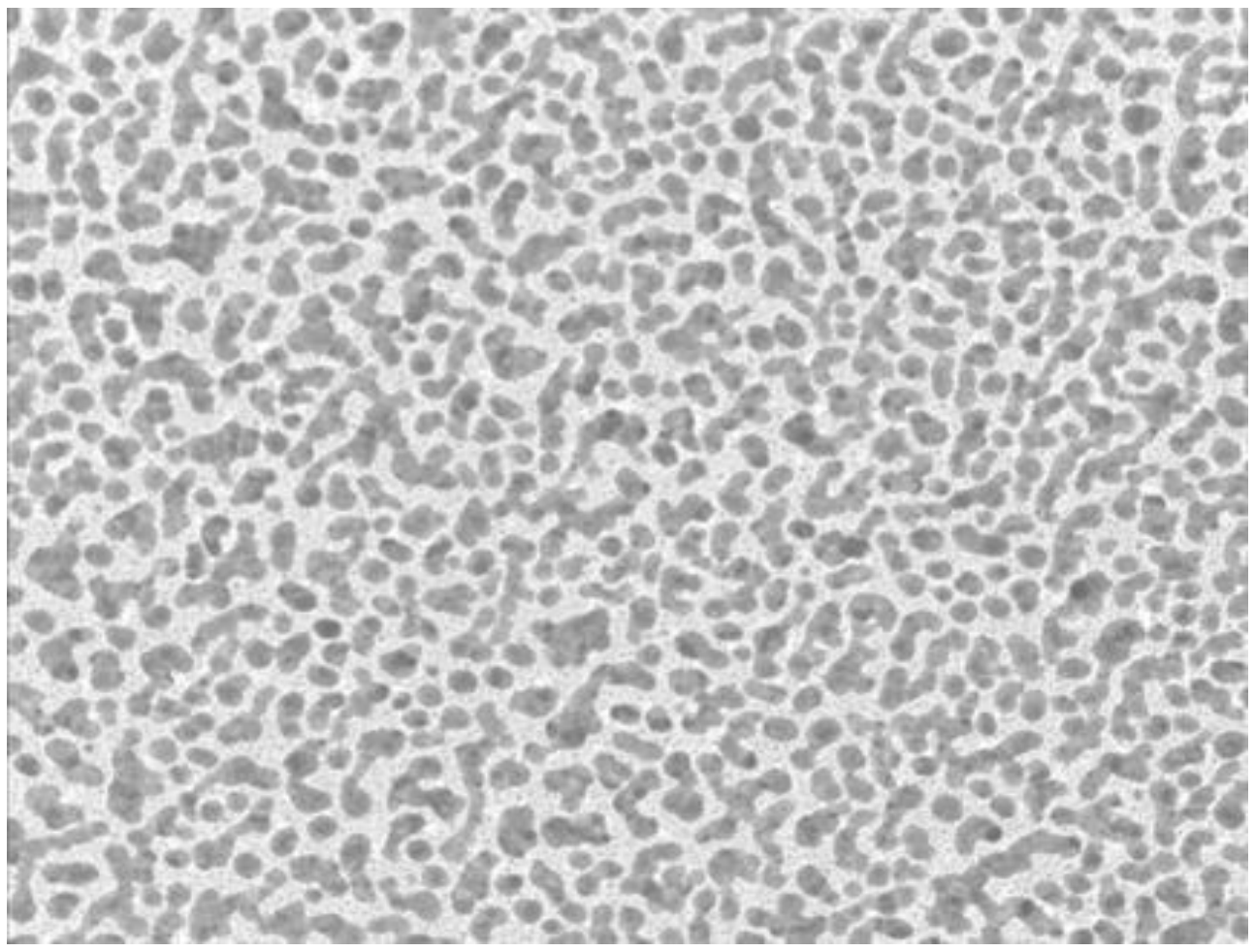

Figure 1. TEM image of copper nanoislands near the percolation limit deposited on a formvar-covered TEM grid. This micrograph features a deposit of 5000 laser pulses. The image field is 93 x 81 nm. See Ref. *(18)*.

As an example of a film near the percolation threshold, Fig. 1 features a TEM micrograph of a 5000-pulse thin-film sample produced via laser ablation of copper, on top of which a pattern of random islands is ostensible. With the same 5000 pulses, but now deposited on a glass substrate, Fig. 2 shows the absorbance of that sample. The baseline for absorption measurements was taken as the absorption of the glass slides.

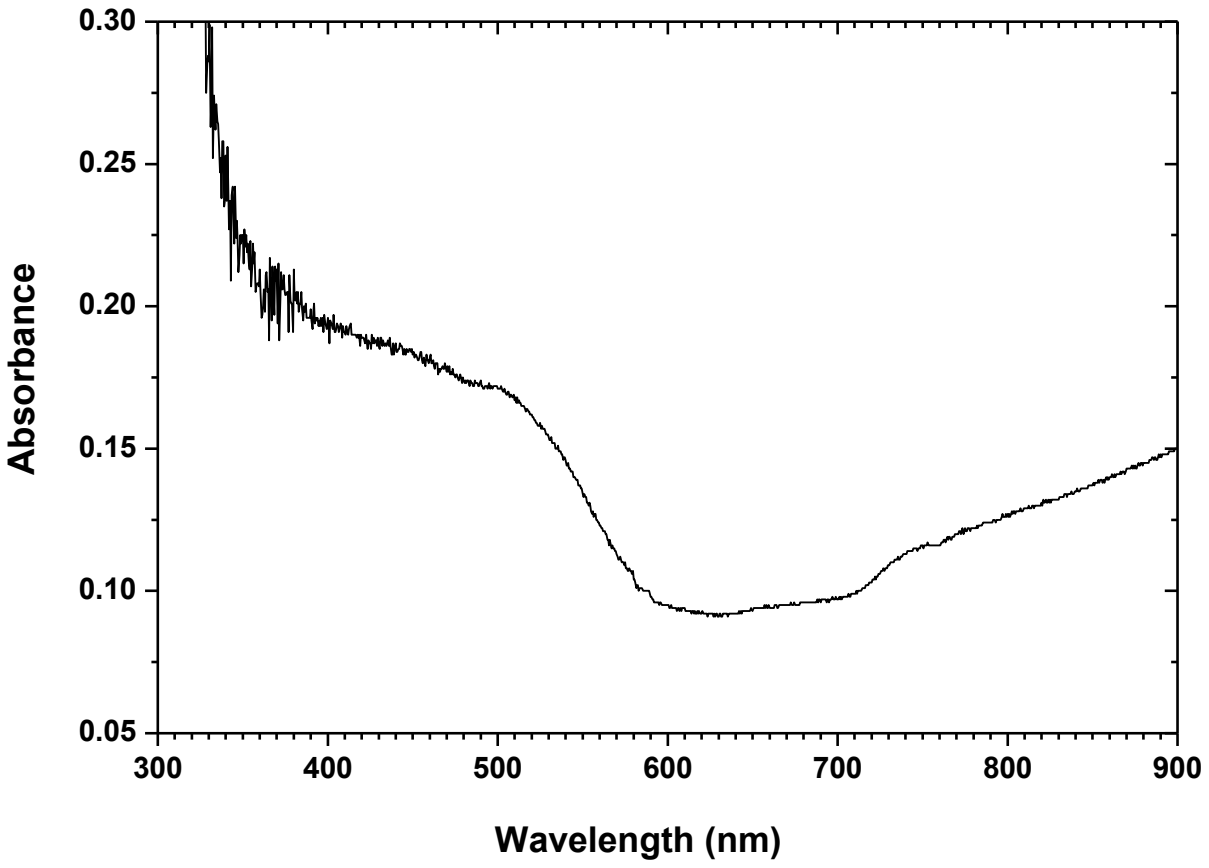

Figure 2. Absorbance spectrum of copper deposited on a glass substrate (5000 laser pulses). Note particularly the wide dip between 600 and 700 nm. The baseline for this measurement is the absorption of the glass slide without any deposited metal.

Already, it can be seen how the absorption *decreases* from its value at 900 nm, reaching its minimum at around 600 nm, and then towards the blue-UV, the absorption rapidly increases again; i.e. once we have added metal to the substrate, the absorption has been reduced around the wavelength at which the surface plasmon resonance of copper is located.

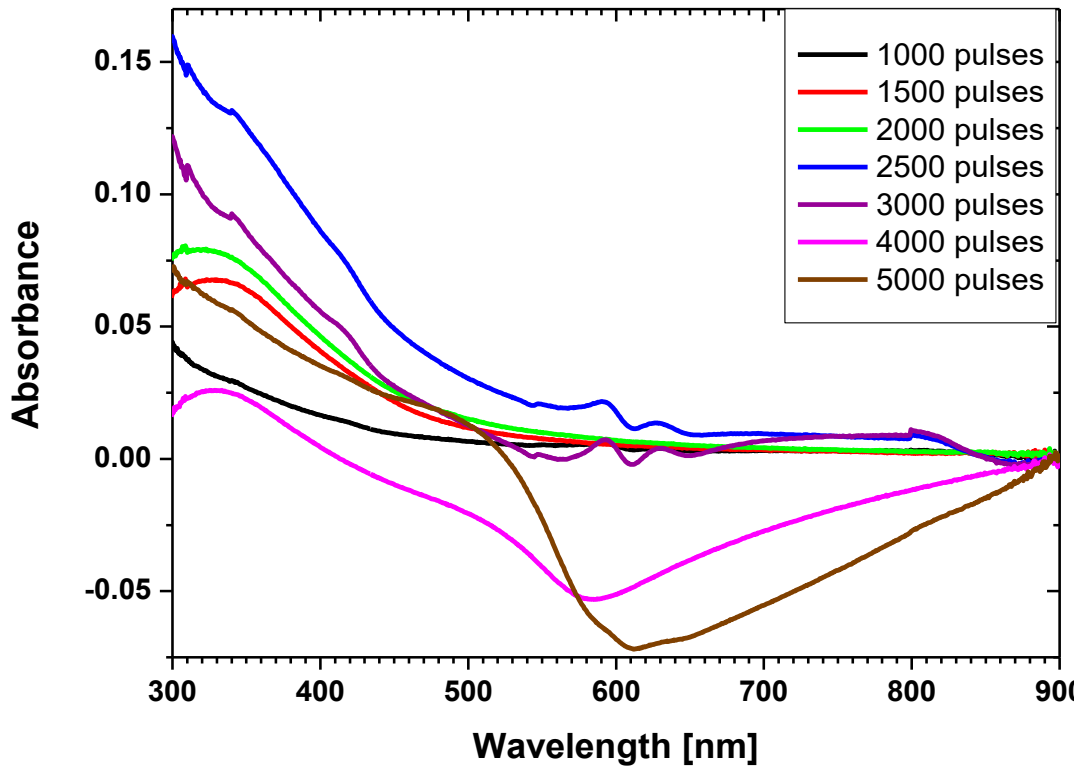

Figure 3. Absorbance of copper films deposited on glass with increasing metal quantity by varying number of laser shots. Note the dip in the copper absorbance, which becomes more pronounced as the number of pulses increases. The small jump at 800 nm is due to a lamp change in the spectrophotometer and a minor un-corrected calibration of the detector.

In Fig. 3, we have plotted the absorbance of a series of copper samples with deposits starting at $10^3$ up to $5x10^3$ laser pulses. For the samples with 1000, 1500 and 2000 laser shots the absorbance is almost flat from 900 to about 450 nm and starts to increase below this wavelength, displaying again a high absorption towards the UV. For the traces of 2500 and 3000 laser shots a double peak structure can be observed. For the sample of 2500 pulses the peaks are located at 590 and 626 nm, whereas for the 3000 pulses film the peaks are found at 592 and 630 nm. It is known that the surface plasmon resonance for copper nanospheres is around 580 nm (19). It is also well known that, when the geometry of the nanoparticles turns elliptic, the single peak splits into two, reflecting the difference of its two semi-axes (see for example Ref. (20). Apart from shape, the dielectric constant of the environment and the proximity to the supporting substrate may change not just the position of the resonances but also their number and width (21). In this case, the double peak structure suggests that, with the metal coverage produced in these samples, although the exact shape of each nanoparticle may not be elliptical, the global optical response of the films is like the one obtained by depositing ellipsoids on a glass substrate. Certainly, as the copper coverage is increased and the percolation limit is nearly reached, where many of the nanoparticles have coalesced, the optical response is likely to be more and more different to the one produced by a separated collection of ellipsoids. In fact, for the samples manufactured with 4000 and 5000 laser shots, we can observe how the absorption decreases (in a similar fashion to the sample used to obtain Figs. 1 and 2, which is near the percolation limit). The minima of absorption for the sample of 4000 laser shots is located at 579.3 nm with a FWHM of about 170 nm, whereas for the sample of 5000 laser shots the minimum occurs at 608 nm with a FWHM of 215 nm. So, for these two last examples, as the copper coverage is increased, the minimum of absorption (maximum of transmission) is red-shifted and the width of the transmission region becomes larger. A similar result has been found in (5) in the context of a Maxwell-Garnet model.

Up to this point, the results displayed in Figs. 2 and 3 show that for copper deposits near the percolation limit, the transmission of light is enhanced near the surface plasmon frequency. Moreover, as the amount of deposited copper increases, the transmission peaks are red shifted and the transmission band broadens.

**Absorption of $TiO_2$/Cu/$TiO_2$ multilayer films.** In a different set of samples, a multilayer structure with different amounts of copper between two layers with constant thickness of $TiO_2$ was prepared. For these samples, an additional $TiO_2$ target was inserted into the ablation chamber. The target-substrate distance was kept also at 0.025 m and the substrates were also glass slides.

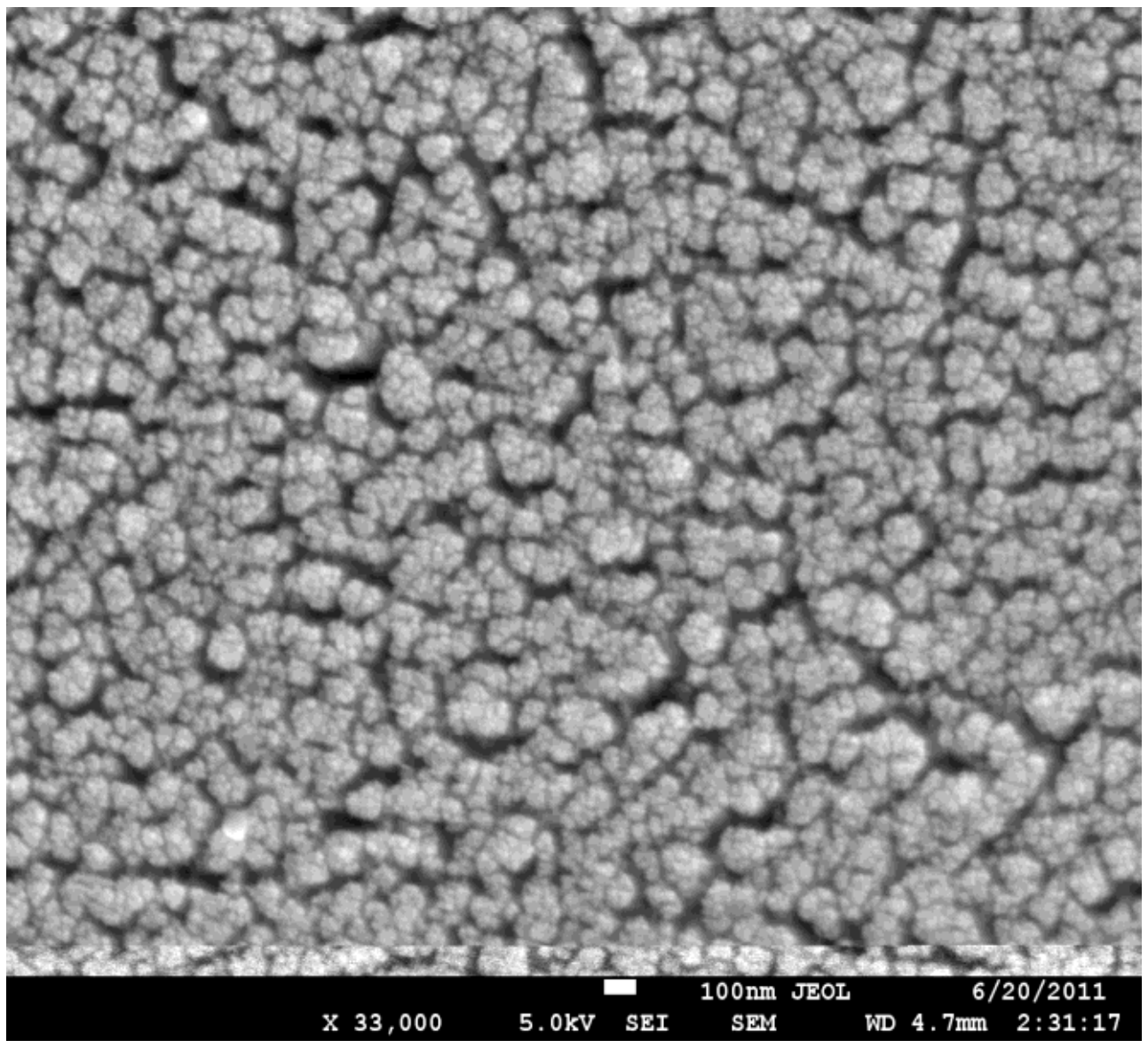

Figure 4. SEM image of a $TiO_2$ film deposited on glass. 9000 laser shots and thickness of 13 nm. The field of the image is 3.6 μm x 3.6 μm.

All the laser ablation procedures for $TiO_2$ were performed at the base pressure of the vacuum chamber, with 9000 laser shots and an energy per pulse of (9.0 ±0.5) mJ. Because the $TiO_2$ layers have been prepared in vacuum, the oxide is not stoichiometric and features a lack of oxygen and, hence, is amorphous. Profilometry and real-time reflectivity measurements show that the thickness of the layers is 13 nm and ellipsometry measurements (LSE Stokes, Gaertner Scientific), indicate that the refractive index is 2.73 at 632.8 nm. Fig. 4 is a SEM image of one of these layers; the deposit shows abundant porosity.

The intermediate copper layers were deposited with the same parameters as in the pure copper samples.

For the absorbance spectra shown in figure 5, the copper deposits started with 2000 laser pulses, then increased by 500 pulses each time until 5000 laser shots. A double peak structure like that in Fig. 2 for the samples with 2000 up to 4500 laser shots in the copper layer can be seen. However, the enhancement of transmission for the traces with 4000 and 5000 pulses easily observed in Fig. 2 has disappeared. For the trace with largest coverage (5000 pulses) the double peak structure of the other samples has been replaced by a very broad feature peaked at ~568 nm.

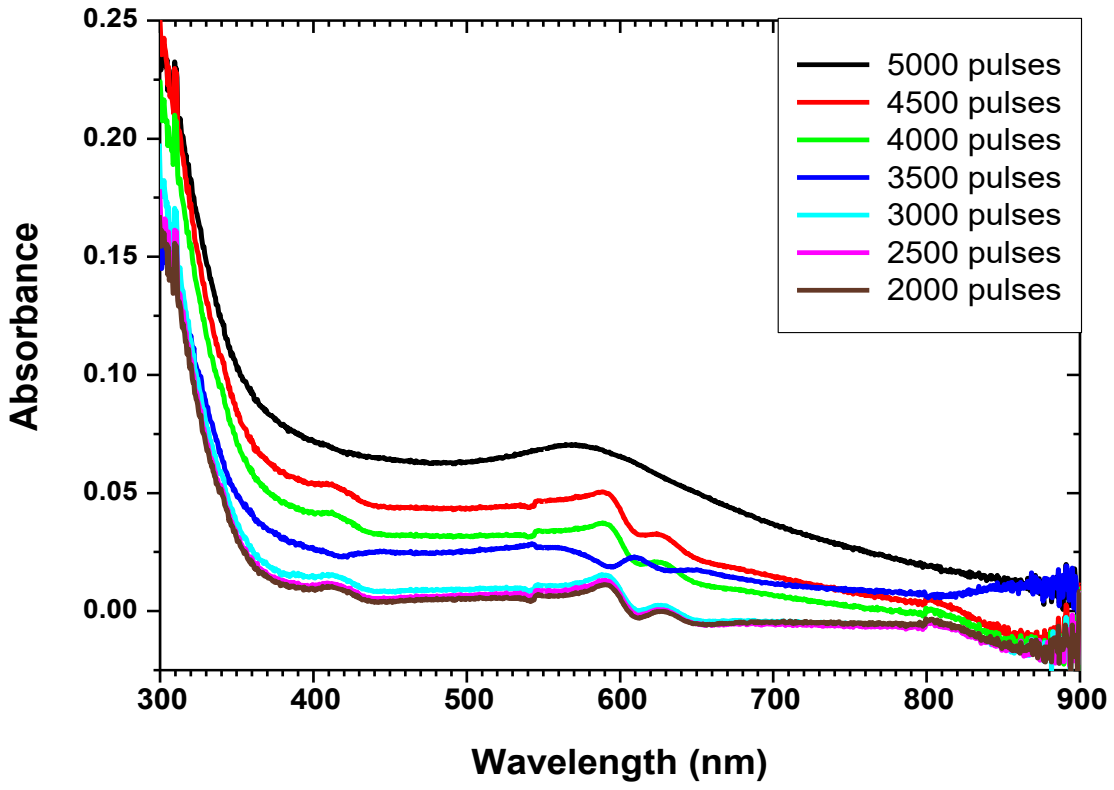

Figure 5. Absorbance spectra of $TiO_2$/Cu/$TiO_2$ multilayers as a function of the number of copper pulses, and constant thickness for the oxide layer (13 nm). $TiO_2$ is amorphous**.**

Hence, in these sandwich-type structures, where the copper is trapped between layers of a high refractive-index medium, the enhanced transmission observed previously is absent, suggesting that this kind of films are useful as sensors for changes in refractive index, as on top of a metallic film another substance may grow or be deposited, changing the enhanced transmission properties of the copper layer.

**Numerical calculations using the discrete dipole approximation (DDA) method and comparison with experimental results.**

To further the understanding of these experiments, light scattering by arbitrary targets has been calculated using open-source software DDSCAT via the discrete dipole approximation; in this method, absorption and scattering off an arbitrary (continuous) target is calculated by dividing it into smaller polarizable dipoles, which interact with one another and with an incident beam of monochromatic light. Dipoles are assumed to occupy positions in a simple cubic lattice; the dielectric properties and the polarization response of the dipoles are linked via the Clausius-Mossotti equation, which holds exactly for a cubic lattice. The resulting scattering problem can be stated analytically in matrix form, and its solution is found numerically (22; 23).

The discrete dipole formalism has been extended to consider periodic targets (customarily known as target unit-cells, TUC's) along directions y and z (light is assumed to propagate along the x-axis) (24). Under these assumptions, the transmission, reflection and absorption coefficients of unpolarized light are given in terms of $S_{11}$ element of the Mueller matrix by the following formulas:

$$T = \sum_{M,N} S_{11}^{(2d)}(M, N, \theta < 90°)$$

$$R = \sum_{M,N} S_{11}^{(2d)}(M, N, \theta > 90°)$$

$$A = 1 - T - R$$

In the above formulas, θ is the usual polar angle in a spherical coordinate system of the scattering problem by an arbitrary target (23); integers M and N represent the orders of diffraction of the problem. Though the formalism is valid for arbitrary orientations of the target, we will always assume that the plane wave is normal to the target (usually a plane, or particles on a plane).

The experimental and numerical work regarding the scattering of light by periodic (11), quasiperiodic (25) and aperiodic (14) scatterers is ample; however, experiments and simulations of the diffraction/transmission properties of light by a random pattern are relatively scarce (15; 6). In this work, several random geometries were explored to model the light transmission phenomenon through thin films.

A contrasted, black-and-white version of Fig. 1 was used as a starting point for our calculations. For the sake of approximation, each pixel of the image pertaining to a nanoisland has been “interpreted” as a discrete dipole. Different thicknesses (in the form of different number of layers) were employed during the calculations.

Fig. 6 features the results of the numerical calculations for light transmission through a TUC constructed from the image in Fig. 1, and for 1 dipole layer; an absorption peak at 386 nm and a transmission maximum near 580 nm are apparent.

The foregoing results are then compared with experimental results of light absorption by thin copper films as a function of the number of pulses shown in Fig. 3. Most of the absorption curves feature strong absorption at the UV in the range 300 to 350 nm; then around 580 nm a double peak where usually the plasmon wavelength is located, then a quasi-constant absorption behaviour.

As seen in figure 3, when the number of pulses exceeds 3000, a minimum for absorption (or, conversely, a transmission peak) appears; the more pronounced dips of absorption occur for the highest number of pulses. This is at odds with the intuitive expectation that an increase in the amount of material deposited on a substrate implies an increased absorption of light. As mentioned before, a red-shift of the minimum as a function of the number of pulses can also be observed.

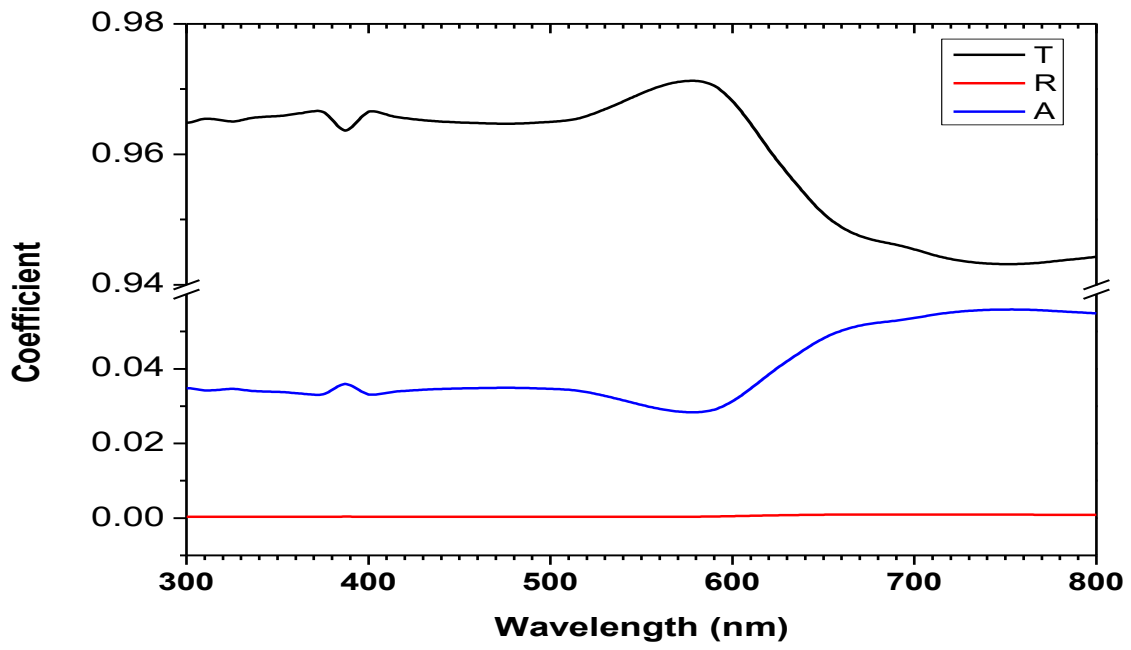

Figure 6. T, R and A coefficients for the TUC from Figure 1. The islands are assumed of copper (index of refraction from Ref. *(26)*); such islands are embedded in vacuum. The target was extended periodically in two dimensions. Thickness of the metal is 1 dipole layer. The minimum of A (maximum of T) is located near 580 nm.

Comparing the graphic shown in Fig. 6 with the trace obtained with 4000 pulses depicted in Fig. 3, we can see that the minima of absorption practically coincide: 573 and 573.3 nm, respectively.

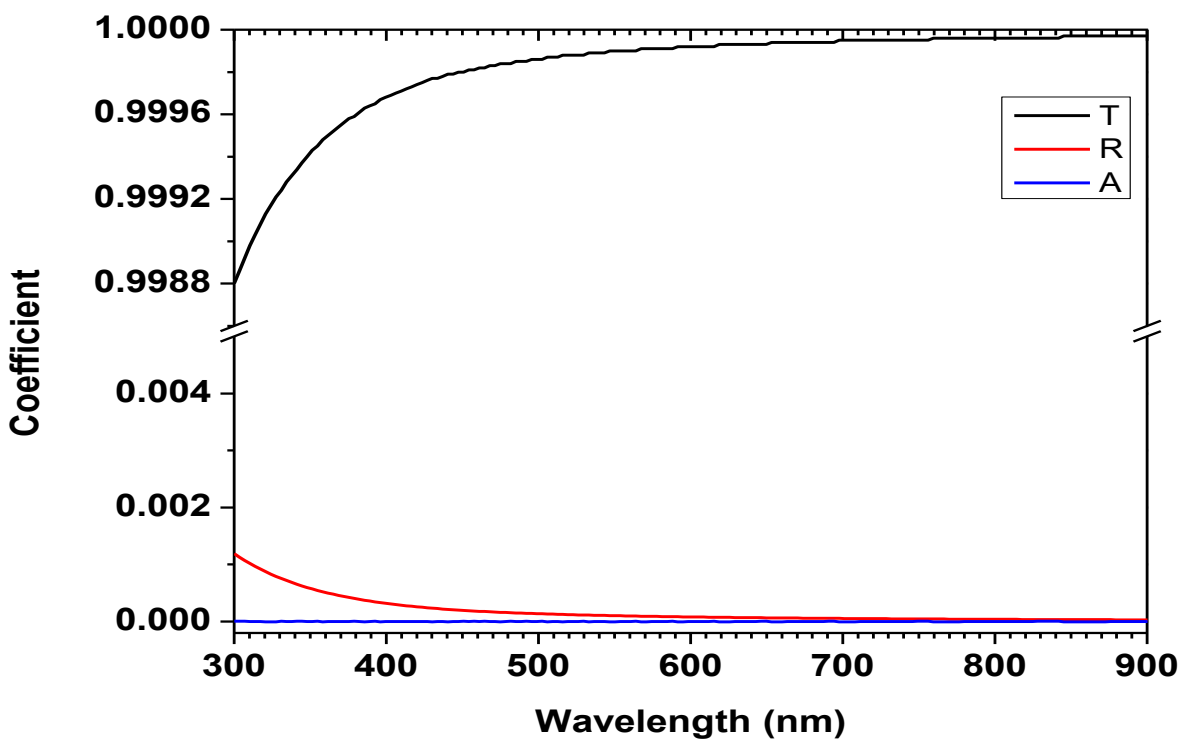

Figure *7*. Transmission, reflection and absorption of light for the same pattern as in Fig. 1, this time assuming islands made of crystalline titanium dioxide ($TiO_2$) (index of refraction from Ref. (27)); such islands are embedded in vacuum. The TUC was extended periodically in two dimensions.

In Fig. 7 we show the results of an analogous calculation of Fig. 6 with a pure dielectric substance of crystalline $TiO_2$. The featureless, monotonic behavior of the curves, as opposed to the rich structure seen for copper is, of course, due to the different properties of a metal film with respect to those of a dielectric film.

As we have seen in Figs. 1 and 2, the light transmission properties of the samples depend on the amount of deposited metal. Simulations were run where the ratio between the area occupied by the islands and the area of the TUC of the sample (henceforth called the occupation ratio) keeping the area of the TUC constant, is changed. The occupation ratio is systematically varied, for occupancies ranging from 10% through 90%; results are shown in Fig. 8. The plasmon peak at ~580 nm redshifts and broadens with increasing occupation rate until ~620 nm.

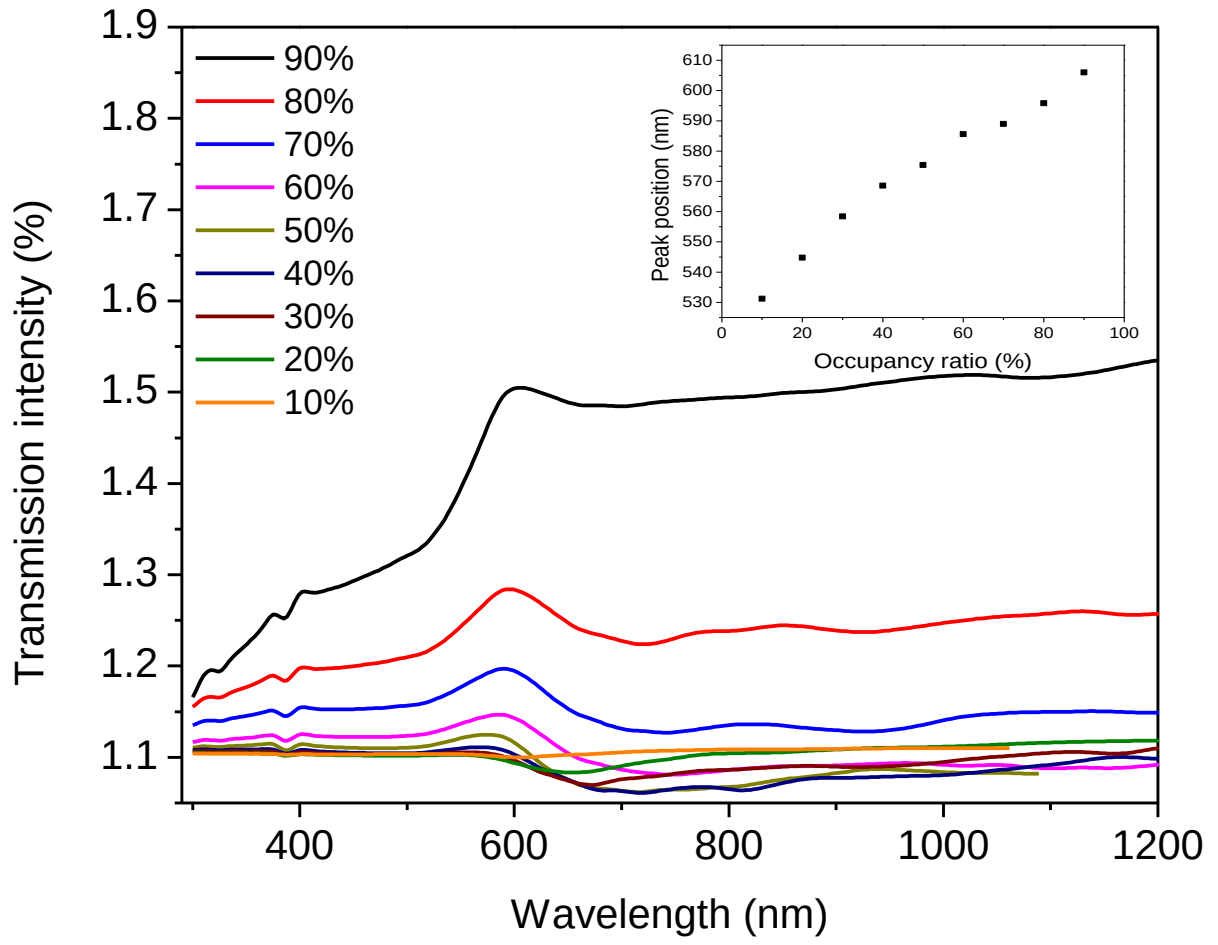

Figure *8*. Transmission curves for 10% through 90% occupation rate. All these curves were calculated assuming a single layer of dipoles. Curves corresponding to 20%-90% have been shifted for clarity of the discussion. Inset: Shift of the surface plasmon polariton position as a function of the occupation ratio; a red-shift occurs as the occupation is increased.

**Possible use as a sensor of changes in the ambient refractive index**

We consider multilayer $TiO_2$-Cu-$TiO_2$ systems to test the effectiveness of these copper films as sensors. As described earlier, Fig. 5 shows the absorbance of a multilayer film where the amorphous oxide layer thickness is kept constant while varying the copper width and, as we can see, the enhanced transmission present in copper island films does not occur here.

Figure 9 shows simulations for a multilayer structure where the copper thickness is changed. The refractive index of the $TiO_2$ layers was taken as the index of the rutile phase (27), while in the experiment the material is amorphous. From this Figure, it is evident that the absorption spectrum of the multilayer structure changes as a function of the thickness of the metallic layer, mirroring this aspect of the experiment (see Fig. 5) even though the crystalline phase of $TiO_2$ was assumed throughout the calculation.

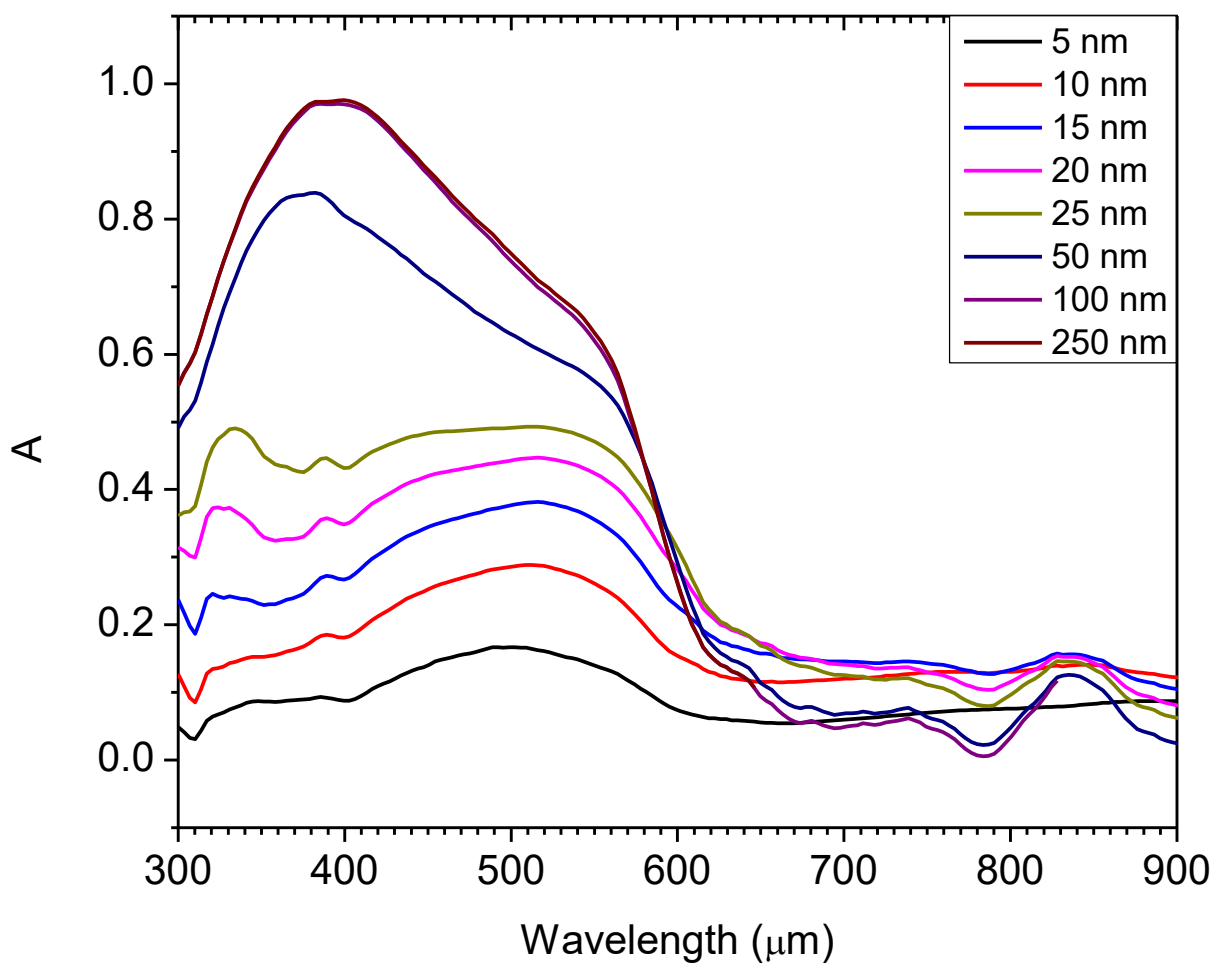

Figure 9. Calculated absorption curves for $TiO_2$-Cu-$TiO_2$ systems as a function of the thickness of the copper layer. A constant thickness of 13 nm for the crystalline $TiO_2$ *(27)* was assumed throughout the calculation.

Also, when the copper-only film is completely embedded in a material with different refractive index, a clear change of the plasmon resonance, in both its amplitude and linewidth, is evident. The light scattered off the pattern in Fig. 1 has been simulated with varying ambient refractive indices and the results can be seen in Fig. 10. A sensor based in a similar idea has been reported in Ref. (29).

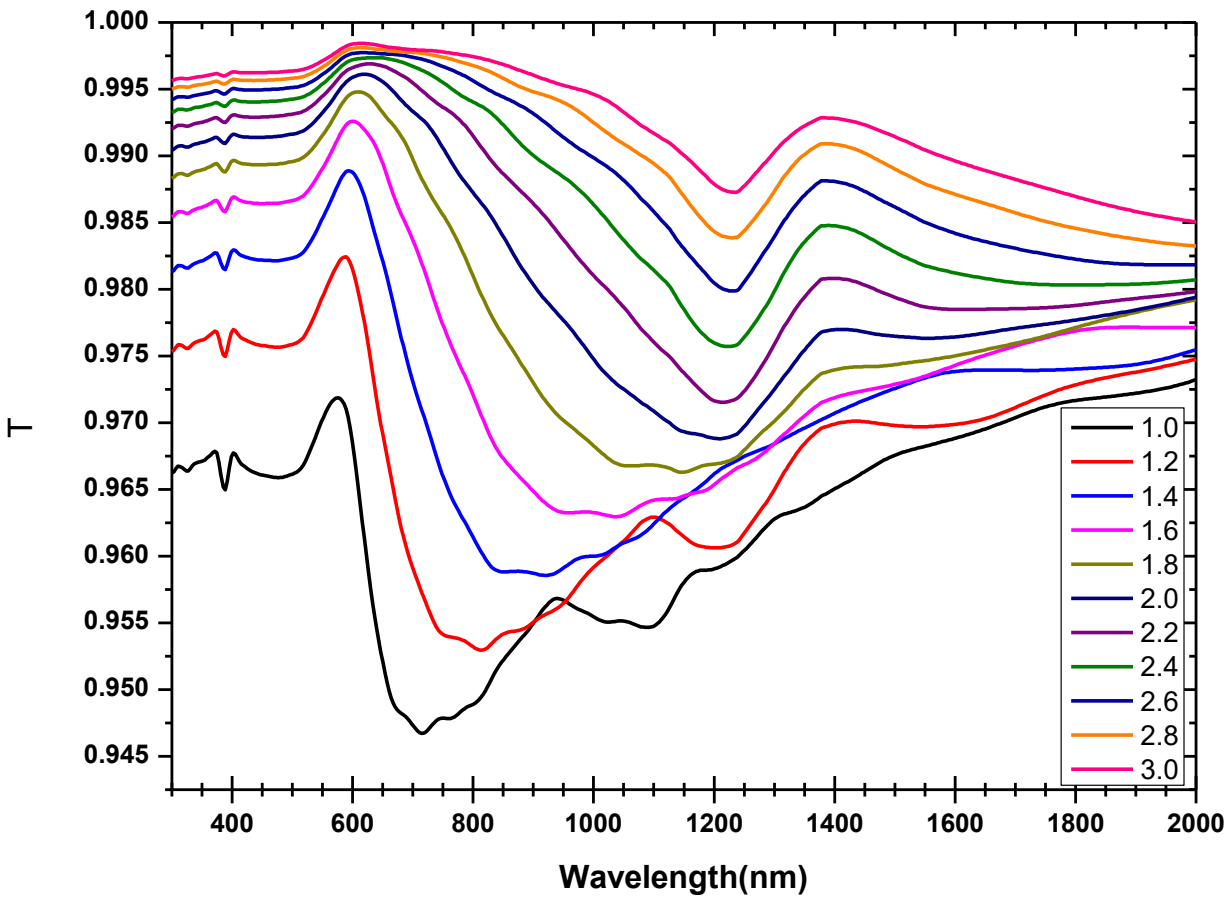

Figure 10. Light transmission as a function of the ambient refractive index for the TUC in Fig. 1.

If we now consider again a TUC as in Fig. 1, made of a single layer of copper dipoles, on top of which lie one, two or three layers of $TiO_2$, we calculate the absorption coefficient of such scattering experiment and show it in Fig. 11. A clear shift in the position of the plasmon peak is noticeable even though the amount of deposited dielectric is relatively minuscule. Such behaviour could be useful if, for example, on top of a quasi-percolated copper film another material, organic or otherwise, grows on or is deposited on top of the metal.

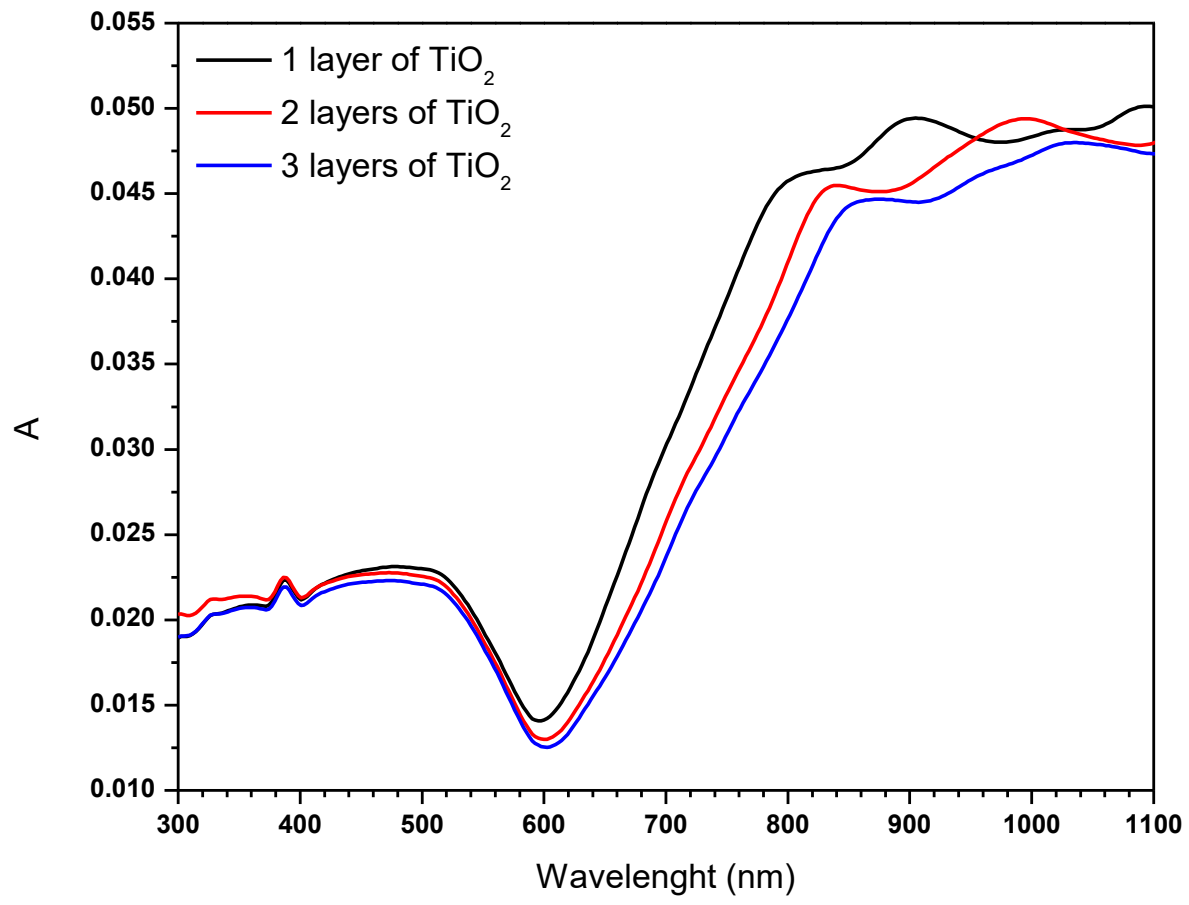

Figure 11. Calculation of the A coefficient using a TUC consisting of a single layer of copper as per Fig. 1, on top of which there are one, two and three uniform layers of $TiO_2$, respectively. The depth and position of the minimum are clearly modified.

In summary, when considering a copper film close to the percolation limit, if the metal is either sandwiched or covered by dielectric materials, the presence of the additional substance can be inferred by monitoring the changes of the position, width and intensity of the plasmon resonance.

The light transmission enhancement exhibited by the copper-only films presented here is reminiscent of the results of extraordinary optical transmission (EOT) in ordered arrays of nanoapertures in metallic films. The results in Ref. (11) can be partially explained via the coupling between incident light and the surface plasmon of the metal involved; an enhancement of the transmission of three orders of magnitude (with respect to the diffraction theory of (10)) was observed. The explanation heavily depended on the periodicity of the lattice.

Such an approach cannot be employed here. Nonetheless, the random TUC pattern from Fig. 1 could be ultimately interpreted as an "irregular" diffraction grating. The dimensions of the scatterers are of sizes less than 10 by 10 nm. The diffraction pattern resulting from the light scattering experiments described in the previous section can be calculated via numerical methods such as the discrete dipole method.

It is interesting to notice that by considering the ratio between transmitted intensity and the space not occupied by the islands, efficiencies ranging between 1.0 and 2.0 are observed in the films presented in this work. On the other hand, in Fig. 2b from (11), transmission efficiencies above 2.0 were observed. However, in the mentioned paper a FIB machine is necessary to fabricate the samples and, while we have used laser ablation in our work, surely just an evaporation chamber would be necessary for manufacturing these island-like films, making the fabrication process much less expensive.

It is also known from percolation theory that for random metal-dielectric composites, in the optical and infrared regions of the spectrum nonlinear processes related to surface plasmon polaritons take place, often in the form of strong local fields (28), so applications like surface-enhanced Raman scattering (SERS) are also plausible.

CONCLUSIONS

We have presented both experimental results and numerical simulations of how, when a thin metallic film approaches the percolation threshold, (i.e. the average separation between islands is smaller than the wavelength of the illuminating light), it shows enhanced transmission of light. We also have shown that this enhancement can be suppressed by placing the metallic film between relatively thick dielectric layers, but also numerical calculations indicate that even few layers of a dielectric on top of the metallic film change the absorption/transmission properties of the samples which, in turn, could find applications for sensor devices.

It is worth pointing out that although EOT in periodic arrays of nanoapertures etched in metallic films, and enhanced optical transmission in random arrays of metallic nanoparticles, are different in several aspects, both phenomena are essentially plasmon-mediated, so it could be possible that a proper model may describe these two apparently dissimilar processes.

METHODS

The experimental setup and general sample preparation are described as follows: copper targets were placed inside a vacuum chamber and irradiated by either the second or the third harmonic of a Q-switched Nd:YAG laser. The vacuum chamber used is a cylindrically symmetric commercial stainless-steel vessel with a volume of approximately 0.01 $m^3$ equipped with several quartz windows (CF flanges) for incoming laser light and diagnostics as well as valves for vacuum pumping and handling diverse gases to be able to control the environment of the ablation process. The chamber is coupled to a turbomolecular pump (Alcatel ATP 150) and fitted with Pirani and cold cathode gauges. The base pressure that can be reached by the system is about $10^{-6}$ mbar.

The laser used for ablation is a Q-switched Nd:YAG Lumonics HY1200 system. In the fundamental wavelength (1064 nm), the system can reach up to 1.2 J per pulse with pulse duration of 10 nanoseconds, the beam diameter is approximately 0.012 m. The laser system is equipped with nonlinear crystals to generate the 2nd (532 nm), 3rd (355 nm) or 4th (266 nm) harmonic. The samples grown for this study were fabricated using 532 or 355 nm depending on which wavelength produced the least sputtering in the samples for a given set of parameters. The repetition rate of the laser can be varied from 1 up to 20 Hz, although during this work it was fixed at 10 Hz.

Before reaching the ablation target, the laser beam is bounced onto two computer-controlled mirrors such that, once an area of the target has been selected, the laser beam will randomly scan it, thus avoiding the drilling of a hole in the copper piece. Moreover, if the laser energy is deposited in just one place, added to the undesired piercing of the target, the splashing caused by the ablation process on the substrate is greatly increased. Evidently, big particulates deposited on the substrate change the optical properties we want to measure; therefore, considerably effort was procured to empirically find the best experimental conditions to avoid splashing. Finally, after the laser beam bounces off the computer-controlled mirror, it traverses a quartz lens – placed outside the vacuum chamber - about 0.5 m from the ablation target. Small displacements of this lens and/or neutral density filters are used to change the laser fluence onto the copper piece.

The cleaning procedure for the copper targets and glass substrates is performed as follows: the copper targets are immersed for about 20 seconds in a 10% diluted solution of a mixture of 75% $HNO_3$, 23% $H_2SO_4$ and 2% HCl, rinsed afterwards twice in distilled water and finally, dried with flowing nitrogen.

The glass substrates are Goldline Microscope slides (Cat. No. 48323-185). They were washed in successive ultrasonic bath solutions containing trichloroethylene, methanol and acetone, submerging them five minutes in each bath solution and then blown with dry air. Inside the vacuum chamber, the distance between the ablation target and the substrate was always kept constant at 0.025 m.

## AUTHOR INFORMATION

### Corresponding Author

* jlhp@xanum.uam.mx

### Author Contributions

The manuscript was written through contributions of all authors.

### Notes

The authors declare no competing interest.

## ACKNOWLEDGMENT

E. M. Rojas-Ruiz thanks the Mexican Consejo Nacional de Ciencia y Tecnología (CONACYT) for the scholarship awarded towards her MSc degree. L. G. Mendoza-Luna is grateful for financial sup-port from PRODEP-SEP. The authors wish to thank: Prof. Ciro Falcony-Guardado (Centro de Investigación y Estudios Avanzados, IPN) for his help in measurements of ellipsometry (LSE Stokes, Gaertner), Prof. J.C. Alonso-Huitrón (IIM-UNAM) for his help in obtaining Fig. 1 (SEM Jeol 7600), Prof. Emanuel Haro-Poniatowski (UAM-I) and Prof. Manuel Fernández-Guasti (UAM-I) for the use of their ablation chamber and Nd:YAG laser, and Prof. Antonio Campero-Celis (UAM-I) for the use of his spectrophotometer (Cary 5E). We also thank the Electron Microscopy Laboratory (Patricia Castillo) at UAM-I. The authors gratefully acknowledge the computing time granted by LANCAD and CONACYT on the supercomputer Yoltla at LSVP UAM-Iztapalapa.

## ABBREVIATIONS

EOT, extraordinary optical transmission; TUC, target unit-cell; DDSCAT, discrete dipole scattering.